\documentclass[letterpaper]{scrartcl}

\usepackage[amsthm]{pamas}

\usepackage{ifthen}

\usepackage{pgflibraryshapes}
\usetikzlibrary{positioning}
\usetikzlibrary{shapes.multipart}

\newcommand{\teq}[1][]{\ifthenelse{\equal{#1}{}}{\mathit{TEQ}}{\mathit{TEQ}(#1)}}
\newcommand{\nteq}{n_{\teq}}
\newcommand{\dom}[1][]{\ifthenelse{\equal{#1}{}}{\overline{D}}{\overline(D)(#1)}}

\newlength{\wordlength}

\title{A tournament of order 24 with two disjoint TEQ-retentive sets}
\author{Felix Brandt \quad and \quad Hans Georg Seedig\\
Technische Universit\"at M\"unchen\\
85748 Munich, Germany\\
\texttt{\small \{brandtf,seedigh\}@in.tum.de}
}

\date{}

\begin{document}

\maketitle

\begin{abstract}
\citet{BCK+11a} have recently disproved a conjecture by \citet{Schw90a} by non-constructively showing the existence of a counterexample with about $10^{136}$ alternatives. We provide a concrete counterexample for Schwartz's conjecture with only $24$ alternatives. 
\end{abstract}

\section{Introduction}

A \emph{tournament $T$} is a pair $(A,{\succ})$, where $A$ is a set of alternatives and~$\succ$ is an asymmetric and complete (and thus irreflexive) binary relation on $A$, usually referred to as the \emph{dominance relation}. 
The dominance relation can be extended to sets of alternatives by writing $X\succ Y$ when $x\succ y$ for all $x\in X$ and $y\in Y$. 
For a tournament $(A,\succ)$, an alternative $x\in A$, and a subset $X \subseteq A$ of alternatives, 
we denote by $\dom_X(x)=\{\,y\in X\mid y \succ x\,\}$ the \emph{dominators} of~$x$.
A \emph{tournament solution} is a function that maps a tournament to a nonempty subset of its alternatives \citep[see, \eg][for further information]{Lasl97a}. Given a tournament $T=(A,\succ)$ and a tournament solution $S$, a nonempty subset of alternatives $X\subseteq A$ is called $S$-\emph{retentive} if $S(\dom_A(x))\subset X$ for all $x \in X$ such that $\dom_A(x)\neq \emptyset$. \citet{Schw90a} defined the tournament equilibrium set ($\teq$) of a given tournament $T=(A,\succ)$ recursively as the union of all inclusion-minimal $\teq$-retentive sets in $T$.

Schwartz conjectured that every tournament contains a \emph{unique} inclusion-minimal $\teq$-retentive set, which was later shown to be equivalent to $\teq$ satisfying any one of a number of desirable properties for tournament solutions \citep{LLL93a,Houy09a,Houy09b,BBFH10a,Bran11b,BrHa11a,Bran11a}.
This conjecture was recently disproved by \citet{BCK+11a} who have shown the existence of a counterexample with about $10^{136}$ alternatives using the probabilistic method. Since it was shown that $\teq$ satisfies the above mentioned desirable properties for all tournaments that are smaller than the smallest counterexample to Schwartz's conjecture, the search for smaller counterexamples remains of interest. In fact, the counterexample found by \citet{BCK+11a} is so large that it has no practical consequences whatsoever for $\teq$. 

In this note, we provide a tournament of order $24$ with two disjoint $\teq$-retentive sets. In contrast to the previous counterexample, both $\teq$-retentive sets are of the same order and even isomorphic. On the other hand, the tournament does not constitute a counterexample to a weakening of Schwartz's conjecture by \citet{Bran11b}.

Let $\nteq$ be the greatest natural number $n$ such that all tournaments of order $n$ or less admit a unique minimal $\teq$-retentive set.
Together with earlier results by \citet{BFHM09a}, we now have that
$12 \leq \nteq \leq 23$.

The counterexample is based on new structural insights about tournament solutions that we will explore further in future work.

\section{The Counterexample}

We define a tournament $T=(A\succ)$ with $24$ alternatives that has two disjoint $\teq$-retentive sets $X=\{x_1,\dots,x_{12}\}$ and $Y=\{y_1,\dots,y_{12}\}$ with $A=X\cup Y$. The two induced subtournaments $T|_X$ and $T|_Y$ are isomorphic. 

\begin{figure}[htbp]
    \centering
    \begin{tikzpicture}[
        dom/.style={->, line width=1pt, shorten >=1mm, shorten <=1mm}
        ]
        \node[draw, ellipse split] (T1){$X_1$ \nodepart{lower}$X_2$};
        \node[draw, ellipse split,node distance=30mm, right of=T1] (T2) {$Y_1$ \nodepart{lower} $Y_2$};
        \node[below of=T1, node distance=12mm] {$X$};
        \node[below of=T2, node distance=12mm] {$Y$};
        \draw[dom] (T1.north east) to (T2.south west);
        \draw[dom] (T2.south west) to[bend left=10] (T1.south east);
        \draw[dom] (T1.south east) to (T2.north west);
        \draw[dom] (T2.north west) to[bend right=10] (T1.north east);
    \end{tikzpicture}
    \caption{The structure of the counterexample. The two subtournaments $T|_X$ and $T|_Y$ are isomorphic and of order $12$.
Furthermore, $X_1\succ Y_2$, $X_2\succ Y_1$, $Y_1\succ X_1$, and $Y_2\succ X_2$.
}
    \label{fig:counterexample-structure}
\end{figure}
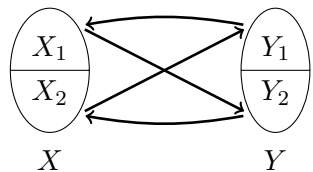

Let $X_1=\{x_1,\ldots,x_6\}$, $X_2=\{x_7,\ldots,x_{12}\}$, $Y_1=\{y_1,\ldots,y_6\}$, and $Y_2=\{y_7,\ldots,y_{12}\}$. Then, the dominance relation between alternatives in $X$ and $Y$ is defined as illustrated in \figref{fig:counterexample-structure}. 

The dominator sets in $T|_X$ (and equivalently in $T|_Y$) are defined as follows:
\[
\begin{array}{lcllcl}
	\dom_X(x_1)		&=&	\{x_4,x_5,x_6,x_8,x_9,x_{12}	\}\text{, }	&
	\dom_X(x_2)		&=&	\{x_1,x_6,x_7,x_{10},x_{12}		\}\text{, }\\	
	\dom_X(x_3)		&=&	\{x_1,x_2,x_6,x_7,x_9,x_{10}	\}\text{, }	&
	\dom_X(x_4)		&=&	\{x_2,x_3,x_7,x_8,x_{11}		\}\text{, }\\
	\dom_X(x_5)		&=&	\{x_2,x_3,x_4,x_8,x_{10},x_{11}	\}\text{, }	&
	\dom_X(x_6)		&=&	\{x_4,x_5,x_9,x_{11},x_{12}		\}\text{, }\\
	\dom_X(x_7)		&=&	\{x_1,x_5,x_6,x_{11},x_{12}		\}\text{, }	&
	\dom_X(x_8)		&=&	\{x_2,x_3,x_6,x_7,x_{12}		\}\text{, }\\	
	 \dom_X(x_9)	&=&	\{x_2,x_4,x_5,x_7,x_8			\}\text{, }	&
	 \dom_X(x_{10})	&=&	\{x_1,x_4,x_6,x_7,x_8,x_9		\}\text{, }\\
	 \dom_X(x_{11})	&=&	\{x_1,x_2,x_3,x_8,x_9,x_{10}	\}\text{, and}	&
	 \dom_X(x_{12})	&=&	\{x_3,x_4,x_5,x_9,x_{10},x_{11}	\}\text{.}
\end{array}
\]
A rather tedious check reveals that
\[
\begin{array}{lcllcl}
	\teq(\dom_A(x_1))  	&=& \{x_4,x_8,x_{12}\}    \subset X\text{, } &
	\teq(\dom_A(x_2))  	&=&  \{x_6,x_{10},x_{12}\} \subset X\text{, }\\
	\teq(\dom_A(x_3))  	&=&  \{x_6,x_7,x_9\} 	    \subset X\text{, } &
	\teq(\dom_A(x_4))  	&=&  \{x_2,x_7,x_{11}\}    \subset X\text{, }\\
	\teq(\dom_A(x_5))  	&=&  \{x_2,x_8,x_{10}\}    \subset X\text{, } &
	\teq(\dom_A(x_6))  	&=&  \{x_4,x_9,x_{11}\}    \subset X\text{, }\\
	\teq(\dom_A(x_7))  	&=&  \{x_1,x_5,x_{11}\}    \subset X\text{, } &
	\teq(\dom_A(x_8))  	&=&  \{x_3,x_6,x_{12}\}    \subset X\text{, }\\
	\teq(\dom_A(x_9))  	&=&  \{x_2,x_5,x_{7}\}     \subset X\text{, } &
	\teq(\dom_A(x_{10}))&=& \{x_4,x_6,x_7\} 	    \subset X\text{, }\\
	\teq(\dom_A(x_{11}))&=& \{x_1,x_2,x_8\} 	    \subset X\text{, and } &
	\teq(\dom_A(x_{12}))&=& \{x_3,x_4,x_9\} 	    \subset X\text{.}\\
\end{array}
\]
Hence, $X$ is $\teq$-retentive in $T$. Moreover, it can be checked that $\teq(\dom_A(y_i))\subset Y$ for all $i\in\{1,\dots,12\}$, which implies that $Y$ is $\teq$-retentive as well. In fact, we even have that, for all $i,j\in\{1,\dots,12\}$, $y_j\in \teq(\dom_A(y_i))$ if and only if $x_j\in \teq(\dom_A(x_i))$.

\end{document}